\documentclass[conference]{IEEEtran}
\def\BibTeX{{\rm B\kern-.05em{\sc i\kern-.025em b}\kern-.08em
    T\kern-.1667em\lower.7ex\hbox{E}\kern-.125emX}}
\usepackage{amsmath}
\usepackage{amsthm}
\usepackage{amsfonts}
\usepackage{amssymb}
\usepackage{graphicx}
\usepackage{cite}
\usepackage{subfigure}
\usepackage{balance}
\usepackage{algorithm}
\usepackage{algorithmic}
\usepackage{enumerate}
\usepackage{color}
\usepackage{array}
\usepackage{indentfirst}
\usepackage{multirow}
\usepackage{booktabs}
\usepackage{color}
\usepackage{slashbox}
\usepackage{diagbox}
\usepackage{threeparttable}
\usepackage[svgnames,table]{xcolor}
\usepackage{bm}
\usepackage{bbm}
\usepackage{subfigure}
\usepackage{epstopdf}

\makeatletter
\newcommand{\tb}{\mathbf}

\newcommand{\Rmnum}[1]{\expandafter\@slowromancap\romannumeral #1@}
\makeatother

\begin{document}
 	\title{Predictive Beamforming for Integrated Sensing and Communication in Vehicular Networks: A Deep Learning Approach }


\author{Chang Liu$^{\ast}$, Weijie Yuan$^{\ddag}$, Shuangyang Li$^{\ast}$, Xuemeng Liu$^{\S}$, Derrick Wing Kwan Ng$^{\ast}$, and Yonghui Li$^{\S}$ \\
$^{\ast}$University of New South Wales, Sydney, Australia \\ $^{\ddag}$Southern University of Science and Technology, Shenzhen, China \\
$^{\S}$University of Sydney, Sydney, Australia \\
Email: $^{\ast}$\{chang.liu19, shuangyang.li, w.k.ng\}@unsw.edu.au,
$^{\ddag}$yuanwj@sustech.edu.cn, \\
$^{\S}$\{xuemeng.liu, yonghui.li\}@sydney.edu.au
}


%
%


\maketitle

\begin{abstract}
The implementation of integrated sensing and communication (ISAC) highly depends on the effective beamforming design exploiting accurate instantaneous channel state information (ICSI). However, channel tracking in ISAC requires large amount of training overhead and prohibitively large computational complexity.
To address this problem, in this paper, we focus on ISAC-assisted vehicular networks and exploit a deep learning approach to implicitly learn the features of historical channels and directly predict the beamforming matrix for the next time slot to maximize the average achievable sum-rate of system, thus bypassing the need of explicit channel tracking for reducing the system signaling overhead.
To this end, a general sum-rate maximization problem with Cramer-Rao lower bounds-based sensing constraints is first formulated for the considered ISAC system.
Then, a historical channels-based convolutional long short-term memory network is designed for predictive beamforming that can exploit the spatial and temporal dependencies of communication channels to further improve the learning performance.
Finally, simulation results show that the proposed method can satisfy the requirement of sensing performance, while its achievable sum-rate can approach the upper bound obtained by a genie-aided scheme with perfect ICSI available.
\end{abstract}


\section{Introduction}
Recently, integrated sensing and communication (ISAC) \cite{liu2021survey, akan2020internet}, where sensing and communication systems is co-designed to share the same frequency spectrum and hardware, has been proposed as an enabling technology for the sixth-generation (6G) wireless systems to reduce the hardware cost and to further improve the spectral efficiency.
Inspire by its powerful capability in both sensing and communication, ISAC can be applied to a variety of scenarios, such as unmanned aerial
vechile (UAV) communications \cite{9456902, liu2020location}, vehicular networks \cite{li2021novel}, and military communications \cite{8883125}.
One important application of ISAC is the ISAC-assisted vehicle-to-instruction (V2I) communication \cite{li2021novel}.
Specifically, by exploiting ISAC for V2I networks, a high-data rate transmission and a high-resolution localization can be expected on a low-cost hardware platform with less spectral resources.
Thus, significant research efforts towards the ISAC-assisted V2I systems have attracted tremendous attention from both academia and industry \cite{liu2021integrated}.

To unleash its potential, various effective schemes and algorithms have been proposed for the design of ISAC-assisted V2I systems.
For example, \cite{huang2020majorcom} designed a carrier agile phased array radar-based ISAC scheme that adopts index modulation for information transmission, which can guarantee the overall performance of both sensing and communication.
Moreover, \cite{liu2020radar} exploited the millimeter wave (mmWave) technology to facilitate the implementation of ISAC, where an extended Kalman filtering (EKF) scheme was developed to accurately track and predict the kinematic parameters of vehicles to further improve the beam alignment performance.
However, the EKF scheme in \cite{liu2020radar} introduces a large computational overhead, such that it is not suitable for practical implementation.
To address this problem, \cite{yuan2020bayesian} developed a Bayesian predictive beamforming framework via exploiting the message passing technique and proposed a low-overhead prediction scheme for V2I networks.
Note that although the aforementioned methods provided some intermediate solutions to the implementation of ISAC, these methods still adopt a cascaded channel prediction-to-beam alignment processing scheme, which is not a truly joint beamforming design guaranteeing both the sensing and communication performance simultaneously. In addition, this cascaded mechanism still brings additional signaling overhead in practice.
Thus, a more pragmatic and truly joint beamforming design is desired for realizing ISAC-assisted V2I networks. 

\begin{figure*}[t]
  \centering
  \includegraphics[width=0.5\linewidth]{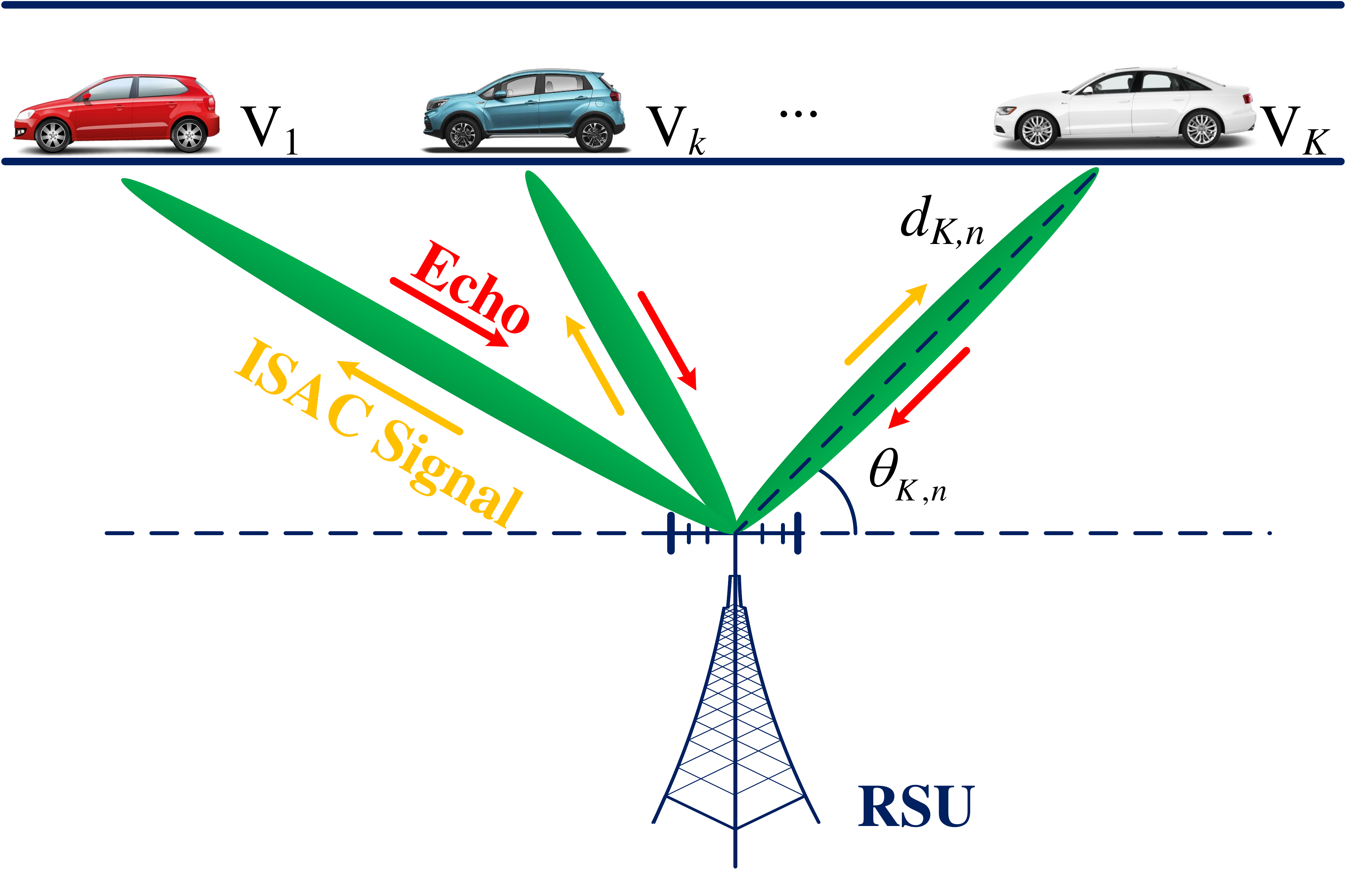}
  \caption{The adopted ISAC-assisted V2I system: An RSU serves $K$ vehicles.}\label{Fig:RSU_scenario}
\end{figure*}

Note that the challenging joint beamforming design is essentially an objective maximization problem which can be well addressed in a data-driven deep learning (DL) approach \cite{goodfellow2016deep, liu2020deepresidual, lxm2020deepresidual, liu2020deeptransfer}.
In this paper, we adopt a DL approach to design a predictive beamforming scheme for ISAC-assisted V2I networks to maximize the average achievable system sum-rate, while guaranteeing the estimation accuracy of motion parameters for vehicles concurrently.
To this end, we first design a communication protocol and derive the Cramer-Rao lower bounds (CRLBs) to characterize the estimation performance accordingly.
Then, a general predictive beamforming problem is formulated to maximize the communication sum-rate while guaranteeing the derived CRLBs-based quality-of-service requirement.
To address the optimization problem, a historical channels-based convolutional long short-term memory (LSTM) network (HCL-Net) is designed, where the historical estimated channels are considered as the input while the convolution and LSTM modules are adopted successively to exploit the spatial and temporal features of communication channels to further improve the learning capability.
In particular, different from existing methods adopting limited motion parameters to facilitate communication task, the developed method can exploit DL technology to intelligently design data-driven features for joint beamforming design taking account of both the sensing and communication performance.
Simulation results show that the proposed scheme can guarantee harsh requirements on the sensing CRLBs and its achievable sum-rate approaches the upper bound obtained by a genie-aided scheme with the availability of instantaneous channel state information (ICSI).

\emph{Notations}:
Superscripts $T$, $H$, and $*$ represent the transpose operation, conjugate transpose operation, and conjugate operation, respectively.
$\mathbb{C}$ and $\mathbb{R}$ denote the sets of complex numbers and real numbers, respectively.
${\mathcal{CN}}( \bm{\mu},\mathbf{\Sigma} )$ represents the circularly symmetric complex Gaussian (CSCG) distribution with the mean vector $\bm{\mu}$ and the covariance matrix $\mathbf{\Sigma}$.
${\mathcal{U}}( a,b )$ represents the uniform distribution within $[a,b]$.
$\tb{0}$ represents a zero vector/matrix.
${\mathbf{1}}_N$ and ${\mathbf{I}}_N$ denote  the $N$-by-$1$ all-ones vector and the $N$-by-$N$ identity matrix, respectively.
$\|\cdot\|$ and $\|\cdot\|_F$ denote the norm of a complex-valued number and the Frobenius norm of a matrix, respectively.
$(\cdot)^{-1}$ is the matrix inverse.
$\mathrm{diag}(\mathbf{x})$ is a diagonal matrix generated based on $\mathbf{x}$ and $\det(\cdot)$ is the determinant of a matrix.
$\frac{\partial f(x,y,\cdots)}{\partial x}$ represents the partial derivative of a function $f(x,y,\cdots)$ with respect to (w.r.t.) variable $x$.
$\mathbb{E}(\cdot)$ is the statistical expectation operation.
$\tb{A} \succeq \mathbf{B}$ means that $\tb{A} - \mathbf{B}$ is positive semidefinite.
$\max(c,d)$ is the maximum between $c$ and $d$.
$\mathrm{Re}\{\cdot\}$ and $\mathrm{Im}\{\cdot\}$ are adopted to denote the real part and the imaginary part of a complex-valued matrix, respectively.

\section{System Model}
As depicted in Fig. \ref{Fig:RSU_scenario}, an ISAC-assisted V2I network is operated in mmWave frequency bands, where a roadside unit (RSU) equipped with a $N_t$-transmit-antenna and $N_r$-receive-antenna uniform linear array (ULA) serves $K$ single-antenna vehicles.
Without special notes, subscripts $k$, $n$, and $t$ refer to the $k$-th, $k \in \{1,2,\cdots,K\}$, vehicle, the $n$-th, $n \in \{1,2,\cdots,N\}$, time slot, and time instant $t$, respectively.
In particular, the full-duplex radio techniques \cite{barneto2021full} is exploited at the RSU to facilitate echoes reception and signal transmission simultaneously \cite{yuan2020bayesian}.

\subsection{Sensing Model}
Denote by $\tb{s}_{n}(t)=[s_{1,n}(t),s_{2,n}(t),\cdots,s_{K,n}(t) ]^T \in \mathbb{C}^{K \times 1}$ the ISAC downlink signal vector, where $s_{k,n}(t)$ is transmit signal for the $k$-th vehicle, denoted by $V_k$.
Let $\tb{W}_n = [\tb{w}_{1,n},\tb{w}_{2,n},\cdots,\tb{w}_{K,n}]$ represent the downlink beamforming matrix with $\tb{w}_{k,n} \in \mathbb{C}^{N_t\times 1}$ being the beamforming vector for $V_k$.
The transmitted signal at the RSU can be formulated as $\tilde{\tb{s}}_n(t) = \tb{W}_n \tb{s}_{n}(t) \in \mathbb{C}^{N_t \times 1}$.
As commonly adopted in e.g., \cite{niu2015survey}, a line-of-sight (LoS) channel model is adopted for the considered mmWave system. Thus, the received echo vector at the RSU can be written as \vspace{-0.1cm}
\begin{equation}\label{r_original}
{\tb{r}}_n(t) =G \sum_{k=1}^K \beta_{k,n} e^{j2\pi \mu_{k,n}t} \tb{b}(\theta_{k,n}) \tb{a}^{\rm H}(\theta_{k,n})\tilde{\tb{s}}_{n}(t-\nu_{k,n}) + \tb{z}(t).
\end{equation}
Here, $G=\sqrt{N_t N_r}$ is the antenna array gain and $\beta_{k,n} = \varrho(2d_{k,n})^{-1}$ is the reflection coefficient with $\varrho$ being the fading coefficient based on the radar cross section and $d_{k,n}$ being the distance between the RSU and $V_k$ at time slot $n$. $\mu_{k,n} \in \mathbb{R}$ and $\nu_{k,n} \in \mathbb{R}$ denote the Doppler frequency and the time-delay, respectively.
$\theta_{k,n}$ denotes the angle between $V_k$ and the RSU.
$\tb{z}(t)\in \mathbb{C}^{N_t\times 1}$ is the CSCG noise vector at the RSU.
In addition, vectors 
\begin{equation}\label{}
  \tb{a}(\theta_{k,n})=\sqrt{\frac{1}{N_t}}[1,e^{-j\pi\cos\theta_{k,n}},\cdots,e^{-j\pi(N_t-1)\cos\theta_{k,n}}]^T
\end{equation} and 
\begin{equation}\label{}
  \tb{b}(\theta_{k,n})=\sqrt{\frac{1}{N_r}}[1,e^{-j\pi\cos\theta_{k,n}},\cdots,e^{-j\pi(N_r-1)\cos\theta_{k,n}}]^T
\end{equation}
denote the transmit and receive steering vectors at the RSU, respectively.
Accordingly, $\mathbf{w}_{k,n}$ is designed based on the predicted angles $\tilde{\theta}_{k,n}$ and can be expressed as $\mathbf{w}_{k,n} = \sqrt{p_{k,n}}\tb{a}(\tilde{\theta}_{k,n})$, where $p_{k,n}$ denotes the corresponding allocated power.
Specifically, we consider a large scale ULA at the RSU with $N_t \gg 1$ and $N_r \gg 1$.
Thus, it is reasonable to assume that the steering vectors for different angles are asymptotically orthogonal and the inter-beam interference between different vehicles is negligible, i.e., $\forall k \neq k^{'}$, we have $|\tb{a}^H(\theta_{k,n})\tb{a}(\theta_{k^{'},n})| \approx 0 $ and $|\tb{b}^H(\theta_{k,n})\tb{b}(\theta_{k^{'},n})| \approx 0$, as commonly adopted in e.g., \cite{liu2020radar, yuan2020bayesian}.
In this case, the RSU can distinguish different vehicles for independent processing via exploiting the angle-of-arrivals of echoes.
Thus, the received echo at the RSU from $V_k$ can be expressed as
\begin{equation}\label{r_kn}
{\tb{r}}_{k,n}(t) = \psi_{k,n} \tb{b}(\theta_{k,n}) \tb{a}^{\rm H}(\theta_{k,n})\tilde{\tb{s}}_{k,n}(t-\nu_{k,n}) + \tb{z}_{k,n}(t),
\end{equation}
where $\psi_{k,n} = G \beta_{k,n} e^{j2\pi \mu_{k,n}t}$ and $\tb{z}_{k,n}(t) \sim \mathcal{CN}(\mathbf{0},\sigma_z^2\mathbf{I}_{N_r})$ is a CSCG vector with $\sigma_z^2$ being the noise variance.

\subsection{Observation Model}
As shown in Fig. \ref{Fig:RSU_scenario}, since the vehicles always keep parallel to the road, the velocity model of can be formulated as
\begin{equation}\label{}
  v_{k,n} = v_{k,n-1} + \Delta v_{k,n-1}, \forall k,n.
\end{equation}
Here, we ignore the direction change of velocity and $v_{k,n}$ denotes the average velocity magnitude of $V_k$ within each time slot duration $\Delta T$. In addition, $\Delta v_{k,n-1}$ is the velocity increment within the time slot $n-1$.
Denote by $v_{\min}$ and $v_{\max}$ the minimum and the maximum of velocity magnitude, respectively, we assume $v_{k,n} \sim \mathcal{U}(v_{\min},v_{\max})$, $\forall k,n$.
To obtain $\tilde{\nu}_{k,n}$ and $\tilde{\mu}_{k,n}$, we can adopt the matched-filtering method, i.e., $\{\tilde{\nu}_{k,n},\tilde{\mu}_{k,n}\} = \arg \max\limits_{\nu,\mu} \left|\int_0^{\Delta T_e}{\tb{r}}_{k,n}(t)s_{k,n}^*(t-\nu)e^{-j2\pi\mu t } dt \right|^2$ where $\Delta T_e \leq \Delta T$ denotes the length of the received echo and $\tilde{\nu}_{k,n}$ and $\tilde{\mu}_{k,n}$ denote the estimated values of $\nu_{k,n}$ and $\mu_{k,n}$, respectively.
In this case, (\ref{r_kn}) can be rewritten as a function of $\theta_{k,n}$, i.e.,
\begin{equation}\label{r_kn_theta}
\begin{aligned}
  \tilde{\tb{r}}_{k,n} & \triangleq \int_0^{\Delta T_e}{\tb{r}}_{k,n}(t)s_{k,n}^*(t-\tilde{\nu}_{k,n})e^{-j2\pi\tilde{\mu}_{k,n} t } dt \\
& =  G \beta_{k,n} \xi \tb{b}(\theta_{k,n}) \tb{a}^{\rm H}(\theta_{k,n})\mathbf{w}_{k,n} + \tilde{\tb{z}}_{k,n},
\end{aligned}
\end{equation}
where $\xi = \hspace{-0.05cm} \int_0^{\Delta T_e}\hspace{-0.05cm}s_{k,n}(t\hspace{-0.03cm}-\hspace{-0.03cm}\nu_{k,n})s_{k,n}^*(t \hspace{-0.02cm}-\hspace{-0.02cm}\tilde{\nu}_{k,n})e^{-j2\pi (\tilde{\mu}_{k,n}t \hspace{-0.02cm}-\hspace{-0.02cm} \mu_{k,n} t)} dt$ represents the matched-filtering gain and $\tilde{\tb{z}}_{k,n} = \int_0^{\Delta T_e} \tb{z}_{k,n}(t)s_{k,n}^*(t-\tilde{\nu}_{k,n})e^{-j2\pi\tilde{\mu}_{k,n} t} dt$ denotes the noise term and $\tilde{\tb{z}}_{k,n} \sim \mathcal{CN}(\mathbf{0},\sigma_r^2\mathbf{I}_{N_r})$ with the noise variance $\sigma_r^2$.
Based on this, the observation models of $d_{k,n}$ and $v_{k,n}$ can be expressed as
\begin{equation}\label{tau_kn_d}
  \tilde{\nu}_{k,n} = \frac{2d_{k,n}}{c} + \epsilon_{k,n}
\end{equation}
and
\begin{equation}\label{mu_kn_v}
  \tilde{\mu}_{k,n} = \frac{2\dot{v}_{k,n} f_c}{c} + \varepsilon_{k,n},
\end{equation}
respectively. Here, $\tilde{\nu}_{k,n}$ and $\tilde{\mu}_{k,n}$ denote the estimates of $\nu_{k,n}$ and $\mu_{k,n}$, respectively. $f_c$ is the carrier frequency, $c$ denotes the speed of signal propagation, and $\dot{v}_{k,n}$ is the radial velocity.
$\epsilon_{k,n} \sim \mathcal{N}(0,\sigma_{\nu}^2)$ and $\varepsilon_{k,n} \sim \mathcal{N}(0,\sigma_{\mu}^2)$ are the estimation errors and $\sigma_{\nu}^2$ and $\sigma_{\mu}^2$ denote the associated noise variances.
In particular, $\sigma_{\nu}^2$ and $\sigma_{\mu}^2$ generally depend on the signal-to-noise ratios ($\mathrm{SNRs}$) at the RSU \cite{liu2020radar, kay1993fundamentals}, and we have $\sigma_{\nu}^2 = \frac{\rho_{\nu}^2\sigma_z^2}{\xi |\psi_{k,n}|^2|\mathbf{a}^H(\theta_{k,n})\mathbf{w}_{k,n}|^2}$ and $\sigma_{\mu}^2 = \frac{\rho_{\mu}^2\sigma_z^2}{\xi |\psi_{k,n}|^2|\mathbf{a}^H(\theta_{k,n})\mathbf{w}_{k,n}|^2}$.

\subsection{Communication Model}
According to Fig. \ref{Fig:RSU_scenario} and the sensing model, the received signal at $V_k$ can be expressed as
\begin{equation}\label{cmodel}
\vartheta_{k,n}(t) = \tilde{G} \alpha_{k,n} e^{j2\pi \mu_{k,n}t}\tb{a}^{ H}(\theta_{k,n})\sum_{k=1}^K \mathbf{w}_{k,n}{s}_{k,n}(t) + \eta_{k,n}(t).
\end{equation}
Here, $\tilde{G} = \sqrt{N_t}$ denotes the antenna gain and $\alpha_{k,n} = \sqrt{ \alpha_0(d_{k,n}/d_0)^{-\zeta}}$ represents the path loss coefficient with $\zeta$ and $\alpha_0$ being the path loss exponent and the path loss at reference distance $d_0$, respectively. Besides, $\eta_{k,n}(t) \sim \mathcal{CN}(0,\sigma_k^2)$ is the noise sample at $V_k$ and $\sigma_k^2$ denotes the associated noise variance.
Based on this, the received signal-to-interference-plus-noise ratio (SINR) at $V_k$ within the time slot $n$ can be expressed as
\begin{equation}\label{SINR}
  \varphi_{k,n}(\mathbf{h}_{k,n},\mathbf{W}_n)
= \frac{ \left| \mathbf{h}_{k,n}^H\mathbf{w}_{k,n}\right|^2 }{\sum_{k^{'} \neq k}^K \left| \mathbf{h}_{k,n}^H\mathbf{w}_{k^{'},n}\right|^2 + \sigma_{k}^2},
\end{equation}
where $\mathbf{h}_{k,n}^H = \tilde{G}\sqrt{\alpha_0(d_{k,n}/d_0)^{-\zeta}}\tb{a}^{H}(\theta_{k,n})$ denotes the effective channel vector between $V_k$ and the RSU.

\begin{figure*}[t]
  \centering
  \includegraphics[width=\linewidth]{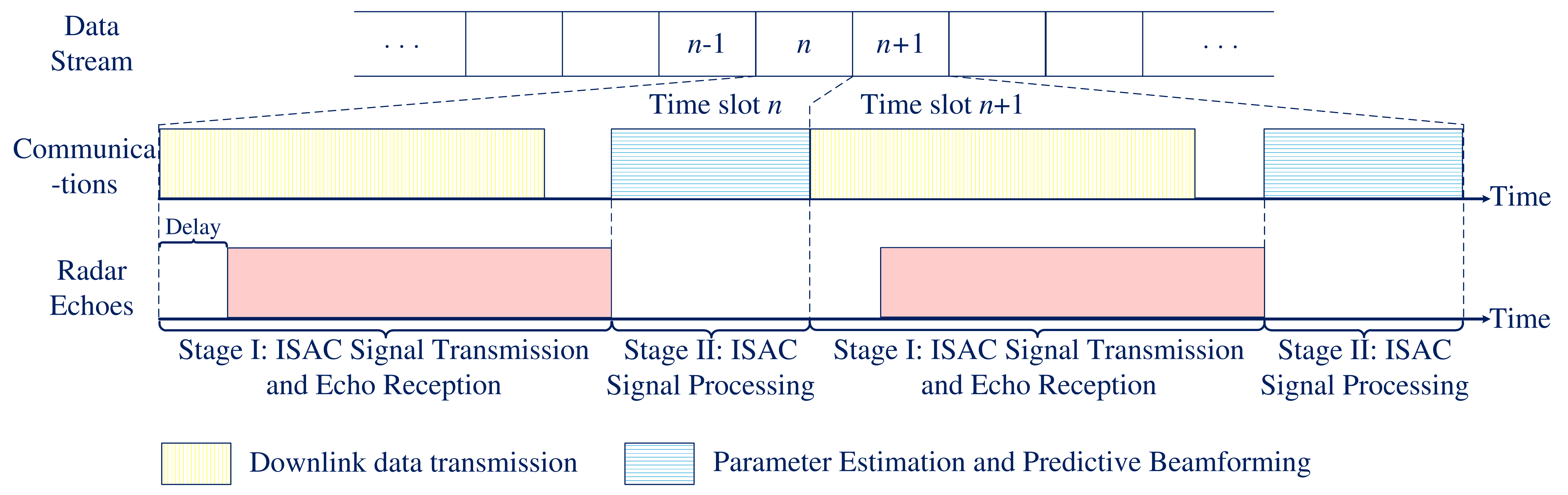}
  \caption{The developed communication protocol for the considered ISAC-assisted V2I system.}\label{Fig:ISAC_frame_structure}
\end{figure*}

\subsection{Proposed Protocol}
In order to reduce the system signaling overhead, we design a predictive transmission protocol for the considered ISAC-assisted V2I system, where $\mathbf{W}_n$ is designed in advance at time slot $n-1$ to bypass the channel tracking/prediction.
As shown in Fig. \ref{Fig:ISAC_frame_structure}, two stages are designed for each time slot in the communication protocol: Stage I is the ISAC signal transmission and echo reception and Stage II is the ISAC signal processing.
In Stage I, the RSU transmits ISAC signals with the optimized beamforming matrix obtained
from the last time slot and receives echoes concurrently via the full-duplex radio technologies \cite{barneto2021full}.
In Stage II, the RSU first estimates vehicles' motion parameters at the current time slot from the received echoes and then predicts the beamforming matrix for the next time slot exploiting the current and historical estimated channels.
This paper mainly focuses on the predictive beamforming design in Stage II, which will be detailed in the next section.

\section{Problem Formulation}
In this section, we aim to formulate the beamforming problem to maximize the average achievable communication sum-rate while maintaining the sensing accuracy for vehicles. In the following, the CRLBs of motion parameter estimation will be first derived to quantify the sensing performance and then we will formulate the optimization problem accordingly.

\subsection{Cramer-Rao Lower Bounds for Parameter Estimation}
Let $\mathbf{y}_{k,n} = [\tilde{\tb{r}}_{k,n}^T, \tilde{\nu}_{k,n}, \tilde{\mu}_{k,n}]^T \in \mathbb{C}^{(N_r + 2) \times 1}$ and $ \mathbf{x}_{k,n} = [\theta_{k,n},d_{k,n},\dot{v}_{k,n}]^T$ represent the observation vector and the motion parameter vector, respectively, we have
\begin{equation}\label{}
  \tb{y}_{k,n} = \mathbf{g}(\tb{x}_{k,n}) + \tb{u}_{k,n},
\end{equation}
where $\mathbf{g}(\cdot)$ is defined in (\ref{r_kn_theta})-(\ref{mu_kn_v}), $\tb{u}_{k,n} = [\tilde{\tb{z}}_{k,n}^T,  \epsilon_{k,n}, \varepsilon_{k,n}]^T$, and $\tb{y}_{k,n} \sim \mathcal{CN}(g(\tb{x}_{k,n}),\mathbf{\Sigma} )$ with $\mathbf{\Sigma} = \mathrm{diag} ([\sigma_r^2\tb{1}_{N_r}, \sigma_{\nu}^2, \sigma_{\mu}^2]) $ being the associated covariance matrix. Thus, the conditional probability density function (PDF) of $\mathbf{y}_{k,n}$ given $\mathbf{x}_{k,n}$ can be expressed as $p(\mathbf{y}_{k,n}|\mathbf{x}_{k,n}) = \frac{1}{\pi^{N_r+2}\det(\mathbf{\Sigma})}\exp\left(-(\mathbf{y}_{k,n} - \mathbf{g}(\mathbf{x}_{k,n})^H)\mathbf{\Sigma}^{-1}(\mathbf{y}_{k,n} - \mathbf{g}(\mathbf{x}_{k,n}))\right)$.
According to \cite{kay1993fundamentals}, the Fisher information matrix (FIM) of $\tb{x}_{k,n}$ is defined as \cite[p. 529]{kay1993fundamentals}
\begin{align}
  \tb{F}(\tb{x}_{k,n}) & = \mathbb{E}\left[\frac{\partial \ln p(\mathbf{y}_{k,n}|\mathbf{x}_{k,n}) }{\partial \mathbf{x}_{k,n}^*} \left(\frac{\partial \ln p(\mathbf{y}_{k,n}|\mathbf{x}_{k,n}) }{\partial \mathbf{x}_{k,n}^*}\right)^H \right] \notag \\
   & = \left(\frac{\partial \tb{g}(\tb{x}_{k,n})}{\partial \mathbf{x}_{k,n}}\right)^H\mathbf{\Sigma}^{-1}\frac{\partial \tb{g}(\tb{x}_{k,n})}{\partial \mathbf{x}_{k,n}},
\end{align}
where
\begin{equation}\label{P}
  \frac{\partial \tb{g}(\mathbf{x}_{k,n})}{\partial \mathbf{x}_{k,n}} = \left[
  \begin{matrix}
  \frac{\partial \tilde{\mathbf{r}}_{k,n}}{\partial \theta_{k,n}}      & 0      & 0      \\
  0      & \frac{2}{c}       & 0      \\
  0      & 0      & \frac{2f_c}{c}      \\
  \end{matrix}
  \right] \in \mathbb{C}^{(N_r+2) \times (N_r+2)}.
\end{equation}
Based on this, the mean squared error (MSE) matrix of $\tb{x}_{k,n}$ is bounded by the following expression:
\begin{equation}\label{}
  \mathbb{E}\left[(\tilde{\tb{x}}_{k,n} - \tb{x}_{k,n})(\tilde{\tb{x}}_{k,n} - \tb{x}_{k,n})^H\right] \succeq \tb{F}^{-1}(\tb{x}_{k,n}).
\end{equation}
Accordingly, we have
\begin{equation}\label{}
  \mathbb{E}\left[(\tilde{\theta}_{k,n} - \theta_{k,n})^2\right] \geq f_{11} \triangleq \mathrm{CRLB}(\theta_{k,n},\tb{w}_{k,n})
\end{equation}
and
\begin{equation}\label{}
  \mathbb{E}\left[(\tilde{d}_{k,n} - d_{k,n})^2\right] \geq f_{22} \triangleq \mathrm{CRLB}(d_{k,n},\tb{w}_{k,n}) ,
\end{equation}
respectively. Here, $f_{ij}$ denotes the $i$-th row and the $j$-th column element of $\tb{F}^{-1}(\tb{x}_{k,n})$ and
\begin{equation}\label{CRLB_theta}
  \mathrm{CRLB}(\theta_{k,n},\tb{w}_{k,n}) =  \left[\frac{1}{\sigma_r^2}\left(\frac{\partial \tilde{\mathbf{r}}_{k,n}}{\partial \theta_{k,n}} \right)^H  \frac{\partial \tilde{\mathbf{r}}_{k,n}}{\partial \theta_{k,n}}\right]^{-1}
\end{equation}
and
\begin{equation}\label{CRLB_d}
  \mathrm{CRLB}(d_{k,n},\tb{w}_{k,n}) =
  \left[\frac{1}{\sigma_{\nu}^2}\left(\frac{2}{c}\right)^2\right]^{-1},
\end{equation}
are the CRLBs of estimating $\theta_{k,n}$ and $d_{k,n}$ given $\tb{w}_{k,n}$, respectively.

\subsection{Problem Formulation}
In this section, we will formulate the predictive beamforming design problem to maximize the average achievable sum-rate via optimizing the beamforming matrix at the RSU subject to the CRLB constraints of the sensing task and the transmit power budget constriant.
Based on the derived CRLBs above, the problem formulation at time slot $n$ can be expressed as
\begin{align}
(\mathrm{P}1):~\mathop{\max}\limits_{{\mathbf{W}}_n } ~ &\mathbb{E}_{\mathbf{H}_{n}|\tb{\Omega}_{n}^{\tau}}
\left\{\sum_{k = 1}^K \log_2\left(1 + \varphi_{k,n}(\mathbf{h}_{k,n},\mathbf{W}_n)\right)  \right\}\label{P1_OP} \\
\mathrm{s.t.}~&
\mathbb{E}_{\mathbf{c}_{n}|\mathbf{\Theta}_{n}^{\tau}} \left\{\frac{1}{K}\sum_{k=1}^K\mathrm{CRLB}(\theta_{k,n},\mathbf{w}_{k,n})\right\} \leq \gamma_{\theta}, \label{P1_C_theta} \\
&
\mathbb{E}_{\mathbf{d}_{n}|\mathbf{D}_{n}^{\tau}} \left\{\frac{1}{K}\sum_{k=1}^K\mathrm{CRLB}(d_{k,n},\mathbf{w}_{k,n})\right\} \leq \gamma_d, \label{P1_C_d} \\
& \|\mathbf{W}_n\|_F^2\leq P. \label{P1_power}
\end{align}
Here, $\mathbb{E}_{\mathbf{H}_{n}|\tb{\Omega}_{n}^{\tau}}\{\cdot\}$ in the objective function denotes the ergodic average w.r.t. $\tb{H}_n = [\tb{h}_{1,n},\tb{h}_{2,n},\cdots,\tb{h}_{K,n}]$, given the historical estimated channels $\tb{\Omega}_{n}^{\tau} \triangleq [\tilde{\mathbf{H}}_{n-1},\tilde{\mathbf{H}}_{n-2},\cdots,\tilde{\mathbf{H}}_{n-\tau}]$,
where $\tilde{\mathbf{H}}_{n} =[\tilde{\mathbf{h}}_{1,n},\tilde{\mathbf{h}}_{2,n},\cdots,\tilde{\mathbf{h}}_{K,n}]$ and $\tilde{\mathbf{h}}_{k,n} = \tilde{G}\sqrt{\alpha_0 (\tilde{d}_{k,n}/{d_0})^{-\zeta}}\tb{a}(\tilde{\theta}_{k,n})$ with $\tilde{\theta}_{k,n}$ and $\tilde{d}_{k,n}$ being the estimated angles and distances, respectively.
Accordingly, $\mathbb{E}_{\mathbf{c}_{n}|\mathbf{\Theta}_{n}^{\tau}}$ denotes the ergodic average w.r.t. $\tb{c}_n = [\theta_{1,n},\theta_{2,n},\cdots,\theta_{K,n}]^T$, given the historical estimated angles $\mathbf{\Theta}_{n}^{\tau} \triangleq [\tilde{\mathbf{c}}_{n-1},\tilde{\mathbf{c}}_{n-2},\cdots,\tilde{\mathbf{c}}_{n-\tau}]$
with $\tilde{\mathbf{c}}_{n} =[\tilde{\theta}_{1,n},\tilde{\theta}_{2,n},\cdots,\tilde{\theta}_{K,n}]^T$.
Similarly, the expectation $\mathbb{E}_{\mathbf{d}_{n}|\mathbf{D}_{n}^{\tau}}$ is taken w.r.t. $\tb{d}_n = [d_{1,n},d_{2,n},\cdots,d_{K,n}]^T$, given the historical estimated distances $\mathbf{D}_{n}^{\tau} \triangleq [\tilde{\mathbf{d}}_{n-1},\tilde{\mathbf{d}}_{n-2},\cdots,\tilde{\mathbf{d}}_{n-\tau}]$
with $\tilde{\mathbf{d}}_{n} =[\tilde{d}_{1,n},\tilde{d}_{2,n},\cdots,\tilde{d}_{K,n}]^T$.
In addition, $\gamma_{\theta}$ and $\gamma_d$ represent the maximum tolerable CRLB thresholds to guarantee the sensing accuracy. $P$ in (\ref{P1_power}) is the power budget at the RSU.
Note that since it is intractable to derive the closed-forms of (\ref{P1_OP}), (\ref{P1_C_theta}), and (\ref{P1_C_d}), solving the problem ($\mathrm{P}1$) is challenging.
Besides, the objective function and constraints are non-convex w.r.t. $\mathbf{w}_{k,n}$, even if the perfect CSI is available.
As an alternative, a data-driven approach can be exploited to design a learning-based predictive beamforming scheme to address problem ($\mathrm{P}1$).

\begin{figure*}[t]
  \centering
  \includegraphics[width=\linewidth]{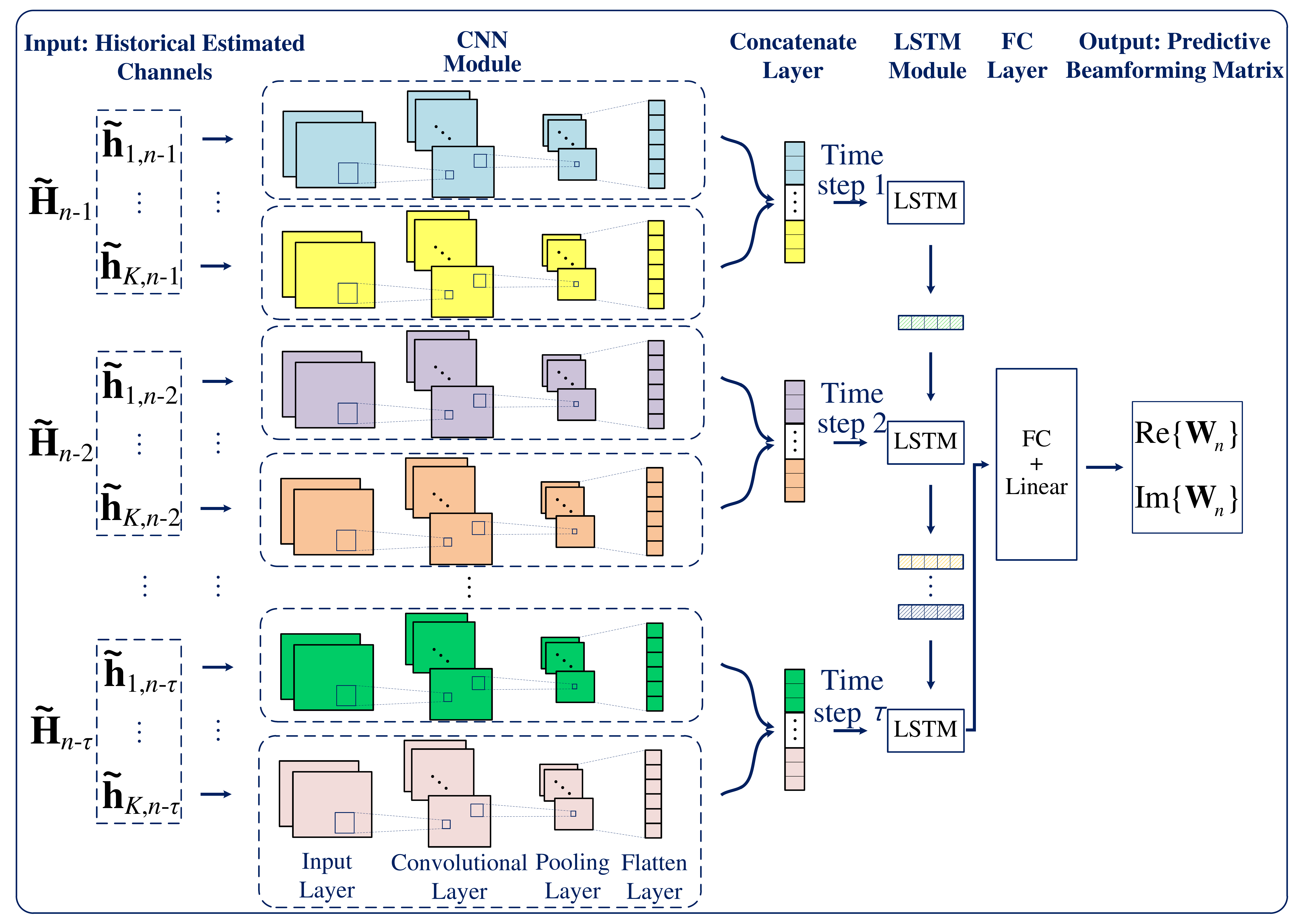}
  \caption{The developed HCL-Net architecture for the predictive beamforming in the considered V2I system.}\label{Fig:HCCL_structure}
\end{figure*}

\section{Proposed HCL-Net-based Predictive Beamforming Design}
Generally, the DL approach is designed to address unconstrained optimization problems \cite{goodfellow2016deep}.
In this case, a penalty method can be exploited to transform the constrained optimization problem (P1) to an unconstrained problem equivalently \cite{gill2019practical}, i.e.,
\begin{equation}\label{P1'}
\begin{aligned}
&(\mathrm{P}1^{'}):~\max\limits_{{\mathbf{W}}_n } ~ \mathbb{E}_{\mathbf{H}_{n}|\tb{\Omega}_{n}^{\tau}}
\left\{\sum_{k = 1}^K\log_2\left(1 + \varphi_{k,n}(\mathbf{h}_{k,n},\mathbf{W}_n)\right)\right\}  \\
& - \lambda_1\max\left(0, \mathbb{E}_{\mathbf{c}_{n}|\mathbf{\Theta}_{n}^{\tau}} \left\{\frac{1}{K}\sum_{k=1}^K\mathrm{CRLB}(\theta_{k,n},\mathbf{w}_{k,n})\right\} - \gamma_{\theta} \right)^2 \\
& - \lambda_2\max\left(0,\mathbb{E}_{\mathbf{d}_{n}|\mathbf{D}_{n}^{\tau}} \left\{\frac{1}{K}\sum_{k=1}^K\mathrm{CRLB}(d_{k,n},\mathbf{w}_{k,n})\right\} - \gamma_d \right)^2 \\
& - \lambda_3\max\left(0, \|\mathbf{W}_n\|_F^2 - P \right)^2,
\end{aligned}
\end{equation}
where $\lambda_\iota \gg 0$, $\iota \in \{1,2,3\}$, is the penalty parameter to control the penalty magnitude.
It should be highlighted that the obtained solution of problem ($\mathrm{P}1^{'}$) can converge to the solution of the original problem (P1) theoretically \cite{gill2019practical}.
In the following, we will first design an HCL-Net to solve problem ($\mathrm{P}1^{'}$) in a data-driven manner and then propose an HCL-Net-based predictive beamforming algorithm accordingly.

\subsection{HCL-Net Architecture}
To better exploit the mapping from the given historical estimated channels to the optimal beamforming matrix, a convolutional LSTM structure \cite{goodfellow2016deep} is adopted to extract the spatial and temporal features of channels to improve the learning ability of the designed HCL-Net which is composed of $K$ convolutional neural network (CNN) modules, one concatenate layer, one LSTM module, and one fully-connected (FC) layer, as illustrated in Fig. \ref{Fig:HCCL_structure}.
The network hyperparameters are presented in Table I accordingly. Here, ``conv.'' is the abbreviation of convolutional and a rectified linear unit (ReLU) is added after each
convolution operation. In addition, a maximization operation is operated for pooling layer and  ``linear'' represents the linear activation function.
To independently exploit the features of real-part and imaginary-part of the input, the complex-valued input is divided into two real-valued parts and the network input can be expressed as
\begin{equation}\label{HCL_input}
  \tilde{\tb{\Omega}}_n^{\tau}= \mathcal{M}([ \mathrm{Re}\{\mathbf{\Omega}_n^{\tau}\}, \mathrm{Im}\{\mathbf{\Omega}_n^{\tau}\} ]),
\end{equation}
where $\mathcal{M}(\cdot)\!:\mathbb{R}^{M \times 2\tau K}\mapsto\mathbb{R}^{\tau \times K \times M \times 2}$ denotes overall the mapping function.
Denote by $h_{\varsigma}(\cdot)$ the expression of HCL-Net with the network parameter $\varsigma$, we have
\begin{equation}\label{HCL_output}
  h_{\varsigma}(\tilde{\tb{\Omega}}_n^{\tau}) = [ \mathrm{Re}\{\mathbf{W}_n\} \mathrm{Im}\{\mathbf{W}_n\} ] \in \mathbb{R}^{K \times M \times 2}.
\end{equation}
Let $\mathcal{F}(\cdot):\mathbb{R}^{K \times M \times 2}\mapsto\mathbb{R}^{K \times M}$ denote the mapping function generating a complex-valued matrix, the optimized beamforming matrix can be expressed as
\begin{equation}\label{W_n_net}
  \mathbf{W}_n = \mathcal{F}(h_{\varsigma}(\tilde{\tb{\Omega}}_n^{\tau})) \in \mathbb{C}^{K \times M}.
\end{equation}

\subsection{HCL-Net-based Predictive Beamforming Algorithm}
The proposed HCL-Net-based predictive beamforming algorithm consists of offline training and online optimization, which will be detailed in the following.

\subsubsection{\underline{Offline Training}}

Given the unlabeled training set $\mathcal{X} = \left\{(\tilde{\mathbf{\Omega}}_k^{\tau(1)},\bar{\mathbf{H}}_{n}^ {(1)} ), (\tilde{\mathbf{\Omega}}_k^{\tau(2)},\bar{\mathbf{H}}_{n}^ {(2)} ), \cdots, \right.  $ $\left. ( \tilde{\mathbf{\Omega}}_k^{\tau(N_e)},\bar{\mathbf{H}}_{n}^{ (N_e)} ) \right\}$, where $N_e$ denotes the number of training examples and $(\tilde{\mathbf{\Omega}}_k^{\tau(i)},\bar{\mathbf{H}}_{n}^ {(i)} )$ is the $i$-th, $i \in \{1,2,\cdots,N_e\}$, training example.
The cost function is then designed to maximize (\ref{P1'}), which is given as
\begin{equation}\label{cost_function}
\begin{aligned}
  &J(\varsigma) =  -\frac{1}{N_e}\sum_{i=1}^{N_e}
\sum_{k = 1}^K\mathrm{log}_2\left( 1 + \frac{\left| (\mathbf{h}_{k,n}^{(i)})^H\mathbf{w}_{k,n}^{(i)}(\varsigma)\right|^2 }{ {\sum_{k^{'} \neq k}^K \left| \mathbf{h}_{k,n}^H\mathbf{w}_{k^{'},n}\right|^2 + \sigma_{k}^2}}\right) \\
& + \lambda_1 \max\left(0,\frac{1}{N_eK}\sum_{i=1}^{N_e} \sum_{k=1}^K\mathrm{CRLB}(\theta_{k,n}^{(i)},\tb{w}_{k,n}^{(i)}(\varsigma)) - \gamma_{\theta} \right)^2 \\
& + \lambda_2\max\left(0, \frac{1}{N_eK}\sum_{i=1}^{N_e} \sum_{k=1}^K\mathrm{CRLB}(d_{k,n}^{(i)},\tb{w}_{k,n}^{(i)}(\varsigma)) - \gamma_d \right)^2 \\
& + \lambda_3 \frac{1}{N_e}\sum_{i=1}^{N_e} \max\left(0, \|\mathbf{W}_{n}^{(i)}(\varsigma)\|_F^2 - P \right)^2,
\end{aligned}
\end{equation}
where $\mathbf{w}_{k,n}^{(i)}(\varsigma)$ is the $k$-th column of $\mathbf{W}_{n}^{(i)}(\varsigma)=\mathcal{F}(h_{\varsigma}(\tilde{\tb{\Omega}}_n^{\tau(i)}))$ in (\ref{W_n_net}).
In addition, a ReLU function $f_{\mathrm{R}}(x) = \max(0,x)$ can be adopted to replace the maximum operators $\max(\cdot,\cdot)$ in (\ref{cost_function}) to facilitate the network training.
Then, based on the training examples, we can adopt a back propagation algorithm (BPA) to minimize (\ref{cost_function}) by progressively updating the network parameters $\varsigma$.
Finally, the optimized predictive beamforming matrix can be expressed as
\begin{equation}\label{well_trained_W_n}
  \mathbf{W}_n^* = \mathcal{F} (h_{\varsigma^*}(\tilde{\tb{\Omega}}_n^{\tau})), \forall n,
\end{equation}
where $h_{\varsigma^*}(\cdot)$ represents the well-trained HCL-Net with the well-trained network parameters $\varsigma^*$ and $\mathcal{F}(\cdot)\hspace{-0.2cm}:\mathbb{R}^{K \times M \times 2}\mapsto\mathbb{R}^{K \times M}$ is defined in (\ref{W_n_net}).

\subsubsection{\underline{Online Beamforming}}
Given a test example $\tb{\Omega}_m^{\tau}$, $m \neq n$, the network input can be expressed as $\tilde{\tb{\Omega}}_m^{\tau}= \mathcal{M}([ \mathrm{Re}\{\mathbf{\Omega}_m^{\tau}\}, \mathrm{Im}\{\mathbf{\Omega}_m^{\tau}\} ])$. Sending $\tilde{\tb{\Omega}}_m^{\tau}$ to the well-trained HCL-Net, we can obtain the optimized predictive beamforming matrix:
\begin{equation}\label{W*_n_net_test}
  \mathbf{W}_m^* = \mathcal{F}(h_{\varsigma^*}(\tilde{\tb{\Omega}}_m^{\tau})).
\end{equation}

\begin{table}[t]
\normalsize
\caption{Hyperparameters of the proposed HCL-Net}\label{Tab:Hyperparameters HCL-Net}\vspace{-0.3cm}
\centering
\small
\renewcommand{\arraystretch}{1.15}
\begin{tabular}{l c}
  \hline \vspace{-0.3cm} \\
   \multicolumn{2}{l}{ \textbf{Input}: $\tilde{\mathbf{\Omega}}_{n}^{\tau}$ with the size of $\tau \times K \times M \times 2$}  \\
  \hline
  \hspace{0.1cm} \textbf{Layers/Modules} & \textbf{Parameters Values}   \\
  \hspace{0.1cm} CNN module - Conv. layer & Size of filters:  $ 4 \times 3 \times 3 \times 2$   \\
  \hspace{0.1cm} CNN module - Pooling layer & Size of filters:  $ 3 \times 3 $   \\
  \hspace{0.1cm} CNN module - Flatten layer & Output shape:  $ 32 \times 1 $   \\
  \hspace{0.1cm} Concatenate layer & Output shape:  $ 96 \times 1 $   \\
  \hspace{0.1cm} LSTM module ($\tau$ time steps) & Size of output:  $ 64 \times 1 $   \\
  \hspace{0.1cm} FC layer & Activation function:  Linear \\
  \hline
   \multicolumn{2}{l}{\textbf{Output}: $[ \mathrm{Re}\{\mathbf{W}_n\}, \mathrm{Im}\{\mathbf{W}_n\} ]$}  \\
  \hline
\end{tabular}
\end{table}

\begin{table}[t]
\small
\centering
\begin{tabular}{l}
\toprule[1.8pt] \vspace{-0.3 cm}\\
\hspace{-0.3cm} \textbf{Algorithm 1} {HCL-Net-based Predictive Beamforming Algorithm} \vspace{0.1 cm} \\
\toprule[1.8pt] \vspace{-0.3 cm}\\
\textbf{Initialization:} $i_t = 0$, $I_t = N_{\max}$, $\varsigma$ with random weights \\
\textbf{Offline Training:} \\
1:\hspace{0.3cm}\textbf{Input:} Training set $\mathcal{X}$\\
2:\hspace{0.6cm}\textbf{while} $i_t \leq I_t $ \textbf{do} \\
3:\hspace{1.1cm}Update $\varsigma$ by BPA to minimize $J(\varsigma)$ in (\ref{cost_function}) \\
\hspace{1.8cm} $i_t = i_t + 1$  \\
4:\hspace{0.6cm}\textbf{end while} \\
5:\hspace{0.3cm}\textbf{Output}:  Well-trained ${h}_{\varsigma^*}( \cdot ) $ as defined in (\ref{well_trained_W_n})\\
\textbf{Online Beamforming:} \\
6:\hspace{0.3cm}\textbf{Input:} Test data  $\tilde{\tb{\Omega}}_m^{\tau}= \mathcal{M}([ \mathrm{Re}\{\mathbf{\Omega}_m^{\tau}\}, \mathrm{Im}\{\mathbf{\Omega}_m^{\tau}\} ])$ \\
7:\hspace{0.6cm}\textbf{do} Predictive Beamforming using\\
\hspace{1.6cm} the well-trained HCL-Net ${h}_{\varsigma^*}( \cdot ) $ \\
8:\hspace{0.3cm}\textbf{Output:} $\mathbf{W}_m^* = \mathcal{F}(h_{\varsigma^*}(\tilde{\tb{\Omega}}_m^{\tau}))$ \vspace{0.2cm}\\
\bottomrule[1.8pt]
\end{tabular}
\vspace{-0.3cm}
\end{table}

\subsubsection{ \underline{Algorithm Steps}}
Based on the above discussion, we can then propose the HCL-Net-based predictive beamforming algorithms, which is summarized in \textbf{Algorithm 1}. Here, for initialization, we set the iteration index and the maximum iteration number as $i_t = 0$ and $I_t = N_{\max}$, respectively.

\section{Simulations and Discussions}
This section will provide simulation results under an ISAC-assisted V2I network operated in mmWave frequency bands.
As introduced in Fig. \ref{Fig:RSU_scenario}, we set $N_t = N_r = 32$, $K = 3$, $\sigma_z^2 = \sigma_k^2 = -80~\mathrm{dBm}$, $\varrho = 10 + 10j$, $\xi= 10$, $\rho_{\nu} = 2.0 \times 10^{-6}$ \cite{yuan2020bayesian} and the RSU is operated with $f_c = 30~\mathrm{GHz}$.
For the path loss model in (\ref{cmodel}), we set $\alpha_0 = -70~\mathrm{dB}$, $d_0 = 1~\mathrm{m}$, and $\zeta = 2.55$ \cite{niu2015survey}.
In particular, a two-dimensional (2D) coordinate system is adopted to describe the locations of the RSU and the $K$ vehicles. Accordingly, the RSU is located at $[0, 0]~\mathrm{(m)}$. Denote by $[\tilde{x}_1,\tilde{y}_1] = [15, 20]~\mathrm{(m)}$, $[\tilde{x}_2,\tilde{y}_2] = [25, 20]~\mathrm{(m)}$, $[\tilde{x}_3,\tilde{y}_3] = [35, 20]~\mathrm{(m)}$, the initial locations of $V_k$ is set as $[x_k,y_k] = [\tilde{x}_k + \Delta x,\tilde{y}_k + \Delta y]~\mathrm{(m)}$ with $\Delta x, \Delta y \sim \mathcal{N}(0,1)$ being random variables to characterize the location uncertainty.
For the motion model of vehicles, we set $\Delta T = 0.02~\mathrm{s}$ and $v_{k,n} \sim \mathcal{U}(8~\mathrm{m/s}, 8.25~\mathrm{m/s})$.
In addition, $\gamma_{\theta} = 0.01~\mathrm{rad}^2$, $\gamma_d = 0.01~\mathrm{m}^2$, and $\lambda_1 = \lambda_2 = \lambda_3 = 10^3$ are adopted for the optimization problem in (\ref{P1'}), where the historical estimated motion parameters are with the normalized MSE of $0.01$.
The adopted hyperparameters for the proposed method are presented in Table \ref{Tab:Hyperparameters HCL-Net} and $N_e = 2,000$ examples are used for offline training.
In addition, three benchmarks are provided for comparisons: (a) upper bound method: Given perfect ICSI, the upper bound performance of problem (P1) can be achieved by a genie-aided scheme, where downlink multiple access interference does not exist and constraints (\ref{P1_C_theta}) and (\ref{P1_C_d}) are ignored; (b) naive DL method in which $\tilde{\theta}_{k,n-1}$ and $\tilde{d}_{k,n-1}$ are adopted as the truth-values at time slot $n$ for beamforming design based on a fully-connected neural network; (c) random beamforming method where the beamforming matrix is set randomly without constraints (\ref{P1_C_theta}) and (\ref{P1_C_d}). Moreover, each point in the simulation results is averaged over $2,000$ Monte Carlo realizations.

\begin{figure}[t]
  \centering
  \includegraphics[width=0.78\linewidth]{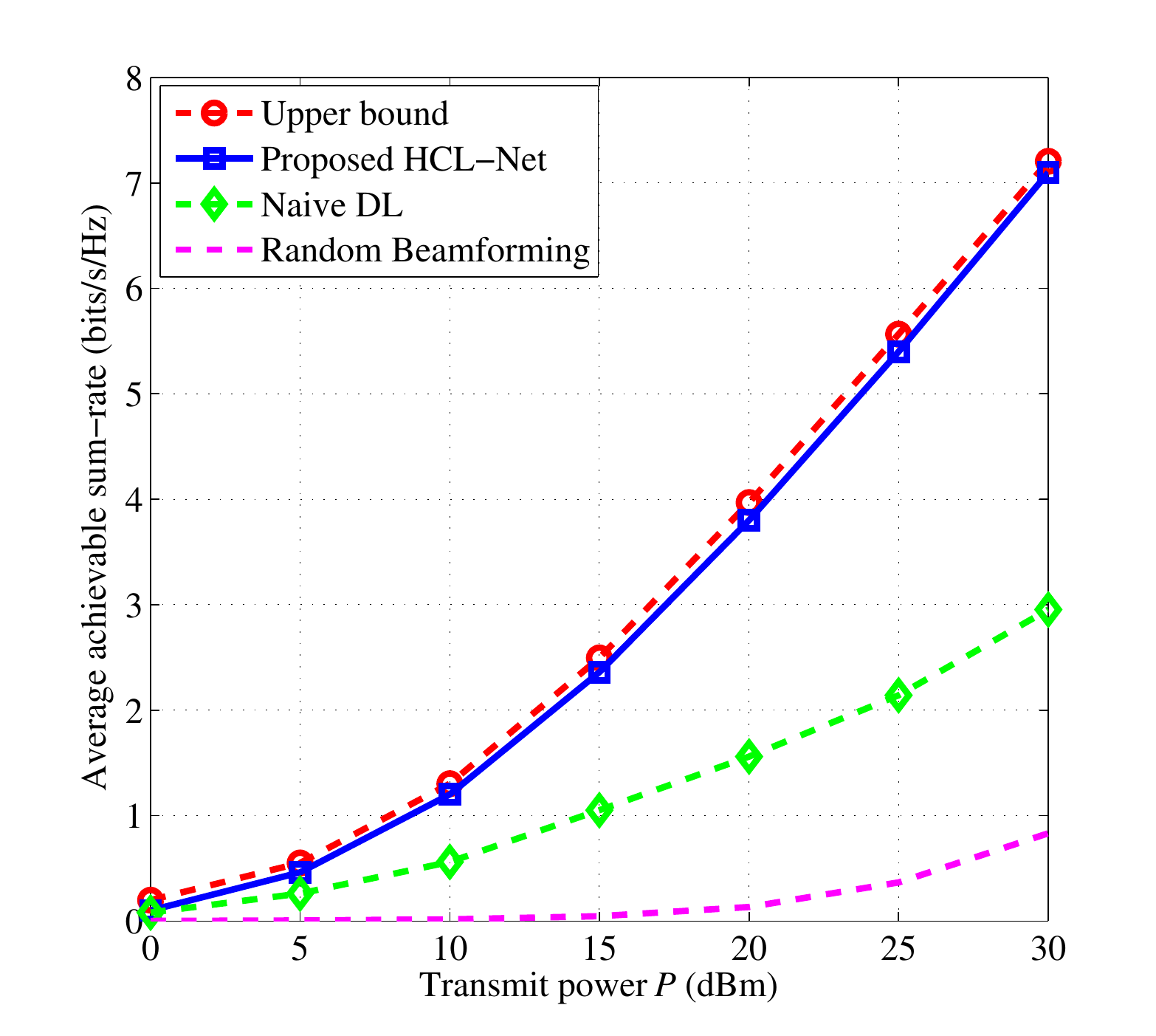}
  \caption{The average achievable sum-rate versus $P$ under $N_t = N_r = 32$.}\label{Fig:Rate_P}
\end{figure}

Fig. \ref{Fig:Rate_P} investigates the average achievable sum-rate versus transmit power budget $P$ at the RSU. It is seen that the random beamforming method performs poorly due to the mismatch between the RSU and vehicles.
Compared with the random beamforming method, the naive DL method achieves a small performance improvement, however, there is still a huge performance gap compared with the upper bound. This is because the naive DL method only exploits the outdated CSI for beamforming and cannot perform a real-time tracking.
In contrast, our proposed method outperforms these two methods significantly and its achievable sum-rate can approach that of the upper bound method exploiting the perfect ICSI.
The reason is that by exploiting features from the historical estimated channels, our proposed method can intelligently predict the beamforming matrix for the next time slot, thus further improving the sum-rate performance.

\begin{figure}[t]
  \centering
  \includegraphics[width=\linewidth]{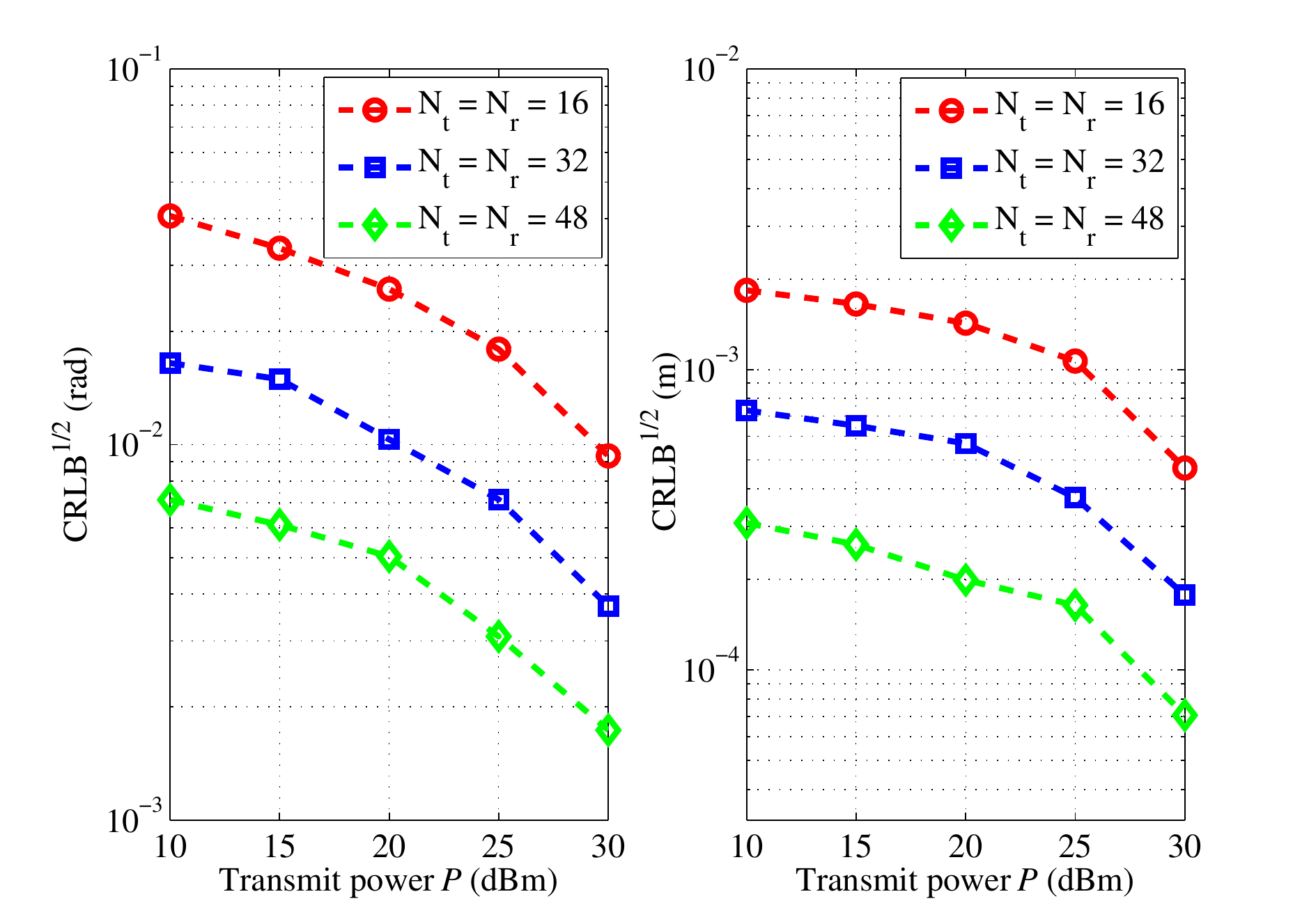}
  \caption{$\mathrm{CRLB}^{1/2}$ in terms of angle and distance estimations.}\label{Fig:Sqrt_CRLB_angle_distance_P}
\end{figure}

To evaluate the sensing performance, Fig. \ref{Fig:Sqrt_CRLB_angle_distance_P} presents the curves of square roots of CRLBs (denoted by $\mathrm{CRLB}^{1/2}$), i.e., the lower bounds of root MSE (RMSE) under different power budgets at the RSU.
It can be observed that the $\mathrm{CRLB}^{1/2}$ values w.r.t. angle estimation (the left hand side of Fig. \ref{Fig:Sqrt_CRLB_angle_distance_P}) and distance estimation (the right hand side of Fig. \ref{Fig:Sqrt_CRLB_angle_distance_P}) are around $10^{-2}~\mathrm{rad}$ and $10^{-3}~\mathrm{m}$, respectively, which are much less than the required maximum thresholds of $\gamma_{\theta}^{1/2} = 0.1~\mathrm{rad}$ and $\gamma_d^{1/2} = 0.1~\mathrm{m}$.
Thus, our proposed method can provide a satisfactory solution to problem (P1).
Also, we can find that the $\mathrm{CRLB}^{1/2}$ can be efficiently reduced by increasing either the transmit power or the number of antennas. These results are expected since the increase of the transmit power or the number of antennas contributes to the improvement of the received SNR at the RSU, thus more accurate angle or distance parameters can be estimated from the received echoes.

\section{Conclusion}
This paper proposed a DL-based predictive beamforming scheme to bypass the explicit channel tracking/prediction to reduce the signaling overhead for the ISAC-assisted V2I networks.
First, a predictive communication protocol was developed and an associated predictive beamforming problem was formulated to maximize the communication sum-rate while guaranteeing the sensing accuracy via the derived CRLB-based constraints.
To solve the formulated problem, a penalty method was exploited to address the problem at hand and an HCL-Net with a convolutional LSTM structure was designed to further improve the learning capability.
Finally, simulation results demonstrated that the proposed method can not only maintain the required sensing accuracy, but also achieves a comparable sum-rate performance to the upper bound obtained by a genie-aided scheme with the perfect ICSI.

\bibliographystyle{ieeetr}

\setlength{\baselineskip}{10pt}

\bibliography{ReferenceSCI2}

\end{document}